\documentstyle[12pt]{article}
\begin{document}

{\Large \bf Two Problems in Classical Mechanics} \\ \\

{\bf Elem\'{e}r E ~Rosinger} \\ \\
{\small \it Department of Mathematics \\ and Applied Mathematics} \\
{\small \it University of Pretoria} \\
{\small \it Pretoria} \\
{\small \it 0002 South Africa} \\
{\small \it eerosinger@hotmail.com} \\ \\

{\bf Abstract} \\

A problem about the present structure of dimensional analysis, and another one about the
differences between solids and fluids are suggested. Both problems appear to have certain
foundational aspects. \\ \\

{\bf 1. Why Scaling, and why the given Groups ?} \\

Dimensional Analysis, Bluman \& Kumei, is one of those fundamental aspects of Classical Physics
which, nevertheless, is often left outside of one's awareness, since it is taken so much for
granted. What it says is that every measurable quantity, say $X$, in Classical Physics has the
{\it dimension}, denoted by $[ X ]$, given by a monomial \\

(1.1) $~~~~~~~ [ X ] ~=~ L^\alpha M^\beta T^\gamma $ \\

where $L$, $M$ and $T$ are, respectively, length, mass and time, and they are the {\it three
Fundamental Mechanical Dimensions}, while $\alpha,~ \beta,~ \gamma \in {\bf R}$ are suitable
exponents. \\
For instance, in the case of velocity $V$, acceleration $A$, or energy $E$, we have,
respectively, $[ V ] = L T^{-1},~ [ A ] = L T^{-2}$ and $[ E ] = L^2 M T^{-2}$. \\

It may, in view of its rather disregarded status, be surprising to see the extent to which the
above concept of dimension is {\it not} trivial. \\

A good example in this respect is the computation by Sir Geoffrey Taylor of the amount of
energy released by the first ever atomic explosion in New Mexico, USA, in the summer of 1944.
All the respective data were, needless to say, classified, and the only information available
to Sir Geoffrey was a motion picture of a few seconds, showing the moment of the explosion and
the consequent expansion of the spherical fireball, see Bluman \& Kumei [pp.9-11]. \\

For a simpler, yet no less surprising example, we recall here a dimensional analysis based
proof of the celebrated theorem of Pythagoras. Let be given the triangle $ABC$

\bigskip
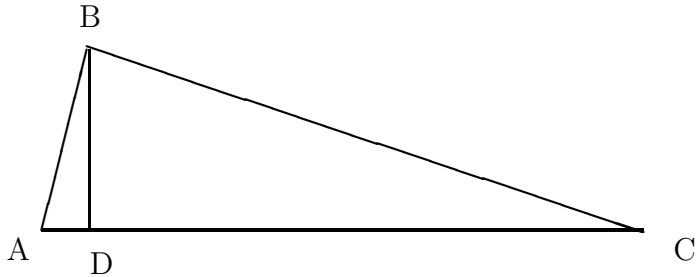
\begin{math}
\setlength{\unitlength}{0.2cm}
\thicklines
\begin{picture}(60,17)

\put(2.7,0){A}
\put(5,2){\line(1,4){3}}
\put(7.5,15.5){B}
\put(5,2){\line(1,0){40}}
\put(8,14.2){\line(3,-1){37}}
\put(47,0){C}
\put(8.2,14){\line(0,-1){12}}
\put(8.2,-0.9){D}

\end{picture}
\end{math} \\

with a right angle at $B$, and let us denote the length of its respective sides by $AB = a,~
BC = b$ and $AC = c$. Further, let us denote by $\psi$ the angle at $A$. \\
Clearly, the area $S(A,B,C)$ of the triangle $ABC$ is perfectly well determined by $c$ and
$\psi$, and it is given by a certain two variable function

$$ S(A,B,C) ~=~ f ( c, \psi ) $$

The point in the above is that, in view of obvious geometric reasons, the function $f$ must be
{\it quadratic} in $c$, namely

$$ f ( c, \psi ) ~=~ c^2~ g ( \psi ) $$

Now, assuming that $AD$ is perpendicular on $AC$, we obtain two right angle triangles $ABD$ and
$BDC$ which are similar with the initial triangle $ABC$. And then adding the areas of the two
smaller triangles, we obtain

$$ S(A,B,C) ~=~ S(A,B,D) + S(B,D,C) $$

thus in terms of the above function $f$, it follows that

$$ f ( c, \psi ) ~=~ f ( a, \psi ) + f ( b, \psi ) $$

which means that

$$ c^2~ g ( \psi ) ~=~ a^2~ g ( \psi ) + b^2~ g ( \psi ) $$

in other words

$$ c^2 ~=~ a^2 ~+~ b^2 $$

that is, the celebrated theorem of Pythagoras. \\

One of the most impressive applications of Dimensional Analysis can be found in the setting up
of a model for three dimensional turbulence by A Kolmogorov, in 1941. \\

Let us return now to the general assumption in Classical Physics formulated in (1.1) above.
Clearly, that relation is equivalent to saying that \\

(1.2) $~~~~~~ [ X ] \in G^3 $ \\

where we denoted by $G^3$ the multiplicative group of all monomials
$ L^\alpha M^\beta T^\gamma$ in (1.1), with the obvious commutative group operation of
multiplication

$$  ( L^\alpha M^\beta T^\gamma )~.~ (  L^{\alpha^\prime} M^{\beta^\prime} T^{\gamma^\prime} )
        ~=~ L^{\alpha + \alpha^\prime} M^{\beta + \beta^\prime} T^{\gamma + \gamma^\prime} $$

In this way, the group $G^3$ is isomorphic with the usual commutative additive group
${\bf R}^3$, according to

$$ G^3 \ni L^\alpha M^\beta T^\gamma  ~~\leftrightarrow~~
                        (\alpha, \beta, \gamma ) \in {\bf R}^3 $$

and the neutral element in $G^3$ is $1 = L^0 M^0 T^0$ which corresponds to the so called {\it
dimensionless} measurable quantities $X$ of Classical Mechanics, namely, for which we have

$$ [ X ] ~=~ 1 $$

The customary reason which is given for the assumption in (1.1) is based on {\it scaling}.
Namely, it is assumed that the {\it units} in which one measures quantities in Classical
Physics are arbitrary and do not influence the mathematical models which express physical laws.
This is why one can simply talk about length, mass and time, and need not specify the
respective units in which they are measured. \\

This, however, clearly conflicts with Quantum Mechanics, where one can no longer consider
arbitrarily small quantities. \\

A second issue related to (1.1) is why precisely those three fundamental mechanical dimensions
of length, mass and time ? Why not other ones ? And if yes, then which other ones ? \\

A third issue one can also raise is the monomial form of the dimensions, as given in (1.1).
After all, with the three fundamental mechanical dimensions $L$, $M$ and $T$, one could as well
construct other groups. \\

In this way, we are led to \\

{\bf Problem 1} \\

Find answers to the above questions. \\ \\

{\bf 2. What are other differences between Fluids and Solids ?} \\

A long recognized way to classify the various {\it states} of matter is to do it according to
{\it two} criteria, Mandelbrot [p. 123], namely

\begin{itemize}

\item flowing versus non-flowing

\item fixed versus variable volume

\end{itemize}

Consequently, we are led to {\it three} possible states. Solids have states which are
non-flowing and with fixed volume. Liquids have states which are flowing and have fixed
volumes. And gases have states which are flowing and with variable volume. \\
The fourth logical possibility, namely, non-flowing and with variable volume is considered not
to be a possible state of usual matter. \\

Here, we shall divide the states of matter only in {\it two} categories, namely

\begin{itemize}

\item solids

\item fluids, which consist of liquids or gases

\end {itemize}

Clearly, therefore, in the above terms, we are led to the following : \\

{\bf First Difference} between solids and fluids : solids are non-flowing, while fluids are
flowing. \\

In Continuum Mechanics one of the long practiced main differences between the mathematical
modelling of solids and fluids is the following. In the respective balance or conservation
equations describing them, the unknowns which model the state of the solid are typically {\it
displacements}, while in the case of fluids are {\it velocities}. Two simple examples
illustrate that difference. \\

The vibrating string, under usual conditions, has the equation

$$ T~ \partial^2_{x x} U ( t, x ) ~=~ m ( x )~ \partial^2_{t t} U ( t, x ) + w ( t, x ),
                            ~~~ t \geq 0,~~~ 0 \leq x \leq L $$

where $L > 0$ is the length of the string placed along the $x$-axis, $U ( t, x )$ is the
lateral {\it displacement} along the perpendicular $y$-axis, $m ( x )$ is the density of the
string at the point $x$, while $w ( t, x )$ is the lateral load at time $t$ and at the point
$x$. \\

On the other hand, the shock wave equation is

$$ \partial_t U ( t, x ) + U ( t, x ) \partial_x U ( t, x ) ~=~ 0,
                                         ~~~ t \geq 0,~~~ x \in {\bf R} $$

where $U ( t, x)$ is the {\it velocity} in the gas at time $t$ and at the point $x$. \\

The usual motivation for this different approach in modelling which prefers displacement in the
case of solids and velocity for fluids is that in fluids displacements can be very large, and
then it is more convenient to consider velocities. \\

In this way we are led to the : \\

{\bf Second difference} between solids and fluids : solids have equations in displacements,
while liquids have equations in velocities. \\

However, it is obvious that solids and fluids can have significantly different properties and
behaviour. And it may appear as quite likely that the respective differences are not taken into
account to a sufficient extent when their usual mathematical modelling is performed. Indeed, in
Continuum Mechanics typically {\it three} kind of relations contribute to the making of the \\

{\bf Usual Mathematical Model} given by

\begin{itemize}

\item balance or conservation equations

\item stress and strain assumptions

\item constitutive relations

\end{itemize}

However, none of these appear to express clearly and significantly enough the {\bf First
Difference} above. Furthermore, one also is lacking a deeper motivation for the {\bf Second
Difference}. \\

In this way, we arrive at the following : \\

{\bf Problem 2} \\

Give a simple and precise mathematical formulation of the difference between solids and fluids
and add it to the {\bf Usual mathematical Model}. \\

{\bf Note} \\

It may turn out that such an augmented mathematical model may be more relevant in the case of
fluids. And to the extent that such would indeed be the case, it may possibly help in a better
modelling of turbulence. \\

\end{document}